\documentclass[12pt,preprint]{aastex}

\shorttitle{Neutron-capture elements in M15}
\shortauthors{Otsuki et al.}

\begin{document}
\title{Neutron-capture elements in the metal-poor globular cluster
M15\footnote{Based on data collected at the Subaru Telescope, which is
operated by the National Astronomical Observatory of Japan.}}

\author{Kaori Otsuki}
\affil{University of Chicago, Chicago, IL 60637, USA}
\email{otsuki@oddjob.uchicago.edu}
\author{Satoshi Honda, Wako Aoki, Toshitaka Kajino}
\affil{National Astronomical Observatory,Mitaka, Tokyo 181-8588 JAPAN}
\email{aoki.wako@nao.ac.jp, honda@optik.mtk.nao.ac.jp, kajino@nao.ac.jp}
\and 
\author{Grant J. Mathews}
\affil{University of Notre Dame, Notre Dame, IN 46556, USA} 
\email{gmathews@nd.edu}

\begin{abstract} 

We report on observations of six giants in the globular cluster M15
(NGC 7078) using the Subaru Telescope to measure neutron-capture
elemental abundances. Our abundance analyses based on high-quality
blue spectra confirm the star-to-star scatter in the abundances of
heavy neutron-capture elements (e.g., Eu), and no significant
s-process contribution to them, as was  found in previous studies. We
have found, for the first time, that there are anti-correlations
between the abundance ratios of light to heavy neutron-capture
elements ([Y/Eu] and [Zr/Eu]) and heavy ones (e.g., Eu).  This
indicates that light neutron-capture elements in these stars cannot be
explained by only a single r-process.  Another process that has
significantly contributed to the light neutron-capture elements is
required to have occurred in M15. Our results suggest a complicated
enrichment history for M15 and its progenitor.

\end{abstract}


\section{Introduction}

The [Fe/H] ratio  for stars
within a globular cluster are generally nearly  identical to within
$\sim$10\% with a few exceptions (e.g., $\omega$- Cen and M22)
\footnote{[A/B] $\equiv \log(N_{\rm
A}/N_{\rm B})- \log(N_{\rm A}/N_{\rm B})_{\odot}$, and $\log
\epsilon{\rm (A)} \equiv \log(N_{\rm A}/N_{\rm H})+12$ for elements A
and B.}.  
Thus,
stars within
a cluster are thought to have been  formed from material having an almost
homogeneous chemical composition.
Furthermore, once globular clusters form, their stars could not have been enriched  
by subsequent epochs of explosive events and star formation. 
Since the  r-process elements observed to be present in 
globular clusters could only have been formed by explosive events,  
their abundance distributions provide 
 a fossil record  of the early Galaxy
enrichment  epoch just prior to the formation of globular clusters.
 
However, \citet{sneden97} have reported a significant spread in
[Ba/Fe] and [Eu/Fe] based on observations of 18 giants in M15,
suggesting a bimodal distribution of these abundances.
\citet{sneden00a} extended the study to 31 stars in M15, although only
the Ba abundances were reported due to the narrower wavelength
coverage.  That study confirmed the scatter of Ba abundances in stars
of M15.  \citet{sneden00b} reported on the abundance patterns of
neutron-capture elements of three red giants in M15 based on
high-resolution blue spectra. They concluded that neutron-capture
elements in M15 are of pure r-process origin based upon their detailed
abundance analysis.  The Ba abundance variations imply that there were
primordial chemical inhomogeneities in the proto-globular cluster in
spite of the uniform Fe abundances.


It has been reported that there is a large scatter in the ratios of
neutron-capture elements to Fe in metal-poor field stars with
[Fe/H]$\leq -2.5$ \citep[e.g.,][]{mcwilliam95,ryan96}.  A scatter also
appears \citep[e.g.,][]{honda04} in the abundance ratios between the
r-process heavy elements (e.g., Ba, Eu, La) and those lighter than Ba
(e.g., Y, Sr, Zr).  This scatter can be interpreted
\citep[e.g.,][]{truran02} as a result of two processes that enrich
neutron-capture elements: one generates both light and heavy r-process
elements, and the other generates only light neutron-capture
elements without increasing the heavy element abundances.  


We have obtained high resolution
blue spectra of seven red giants in M15. These data are of higher
quality than that of previous studies.  
There are two purposes of this study: 
1) to confirm the [Eu/Fe] scatter
in M15 stars which has been reported by Sneden et al.;  and 2) to study the
enrichment of other light and heavy neutron capture elements in M15.
In this Letter we report the distinct result found for abundances of Y,
Zr, La and Eu for six giants in M15, while the results of our full
abundance analyses will be reported in a future publication.

\section{Observations}\label{sec:obs}

We have selected seven red giants in M15 from the list of \citet{sneden97}
to cover stars having high and low abundances of Ba and Eu. 
For comparison purposes
we  also observed the bright halo red giant HD~221170.  This comparison star  
is known to have a moderate enhancement 
of r-process elements \citep{yushchenko05}.

High resolution spectroscopy was obtained with the Subaru Telescope
High Dispersion Spectrograph \citep[HDS; ][]{noguchi02} on July 25 and
26, 2004. The typical seeing resolution was 0.45~arcsec. This excellent
seeing was very important as 
it allowed us to separate the target from surrounding
stars in the cluster. The wavelength range covered was
3550--5250~{\AA}, with a resolving power $R = 50,000$. Five 30 minute
exposures were obtained for K146 and K1040, which are the faintest stars 
at the $B$ magnitude among our sample.  Three exposures (for a total of
90~minutes) were obtained for the other stars.  
HD~221170 was observed with the same setup. Another spectrum of this
object covering 4030--6800~{\AA} with similar quality was obtained in
June 2005 to measure Ba and Eu lines in the red range.

\begin{figure} 
\includegraphics[width=12cm]{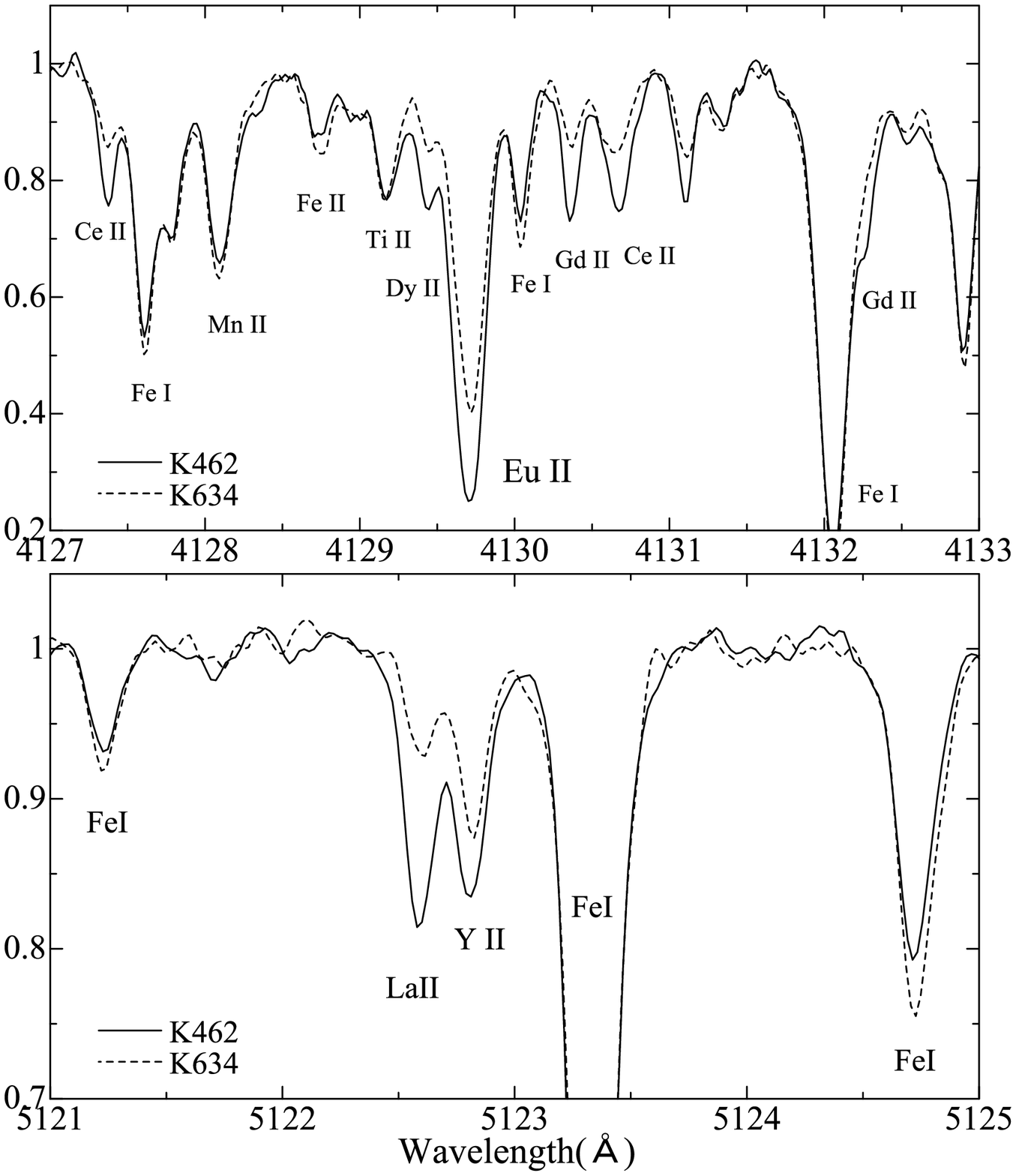} 
\caption[]{Examples of spectra of M15 stars. The upper panel shows
that the absorption features of heavy neutron-capture elements (Ce,
Eu, Gd, and Dy) in K462 are significantly stronger than those in
K634, while the features of Ti, Fe, and Mn are almost identical. The
lower panel demonstrates that the difference between the two stars of the absorption feature
of the light neutron-capture element Y is not as large as that of the
heavy neutron-capture element La.}

\label{fig:sp} 
\end{figure}

Standard data reduction procedures (bias subtraction, flat-fielding,
background subtraction, extraction, and wavelength calibration) were
carried out with the IRAF echelle package\footnote{IRAF is distributed
by the National Optical Astronomy Observatories, which is operated by
the Association of Universities for Research in Astronomy, Inc. under
cooperative agreement with the National Science Foundation.}. The
number of photons collected at 4300~{\AA} for M15 stars ranges from
10,400 to 12,000 per 0.18~km~s$^{-1}$ pixel, indicating that our data
set is quite homogeneous. After the data reduction, one star (K479)
was excluded from our analysis because this star shows significantly
broadened spectral lines.  This suggests a significant contamination
from the light of other stars, even though no signature of this
contamination was found in our slit viewer image nor in the archived
HST image.\footnote{The images are based on
observations made with the NASA/ESA Hubble Space Telescope, obtained
from the data archive at the Space Telescope Science Institute.  STScI
is operated by the Association of Universities for Research in
Astronomy, Inc. under NASA contract NAS 5-26555.}
Figure~\ref{fig:sp} shows a comparison of the spectra of K462 and K634
that have very similar features due to \ion{Fe}{1}, \ion{Fe}{2}, and
\ion{Mn}{2}. However, the absorption features due to neutron-capture
elements are significantly stronger in K462 than in K634. Moreover,
the comparison at 5123~{\AA} (lower panel) demonstrates that the
difference in the La abundance between the two stars is greater than
the difference for Y.

The equivalent widths of neutron-capture elements given in
Table~\ref{tab:ew}, and  also iron, were measured by Gaussian fitting. We
have applied the standard abundance analysis with SPTOOL (Y. Takeda,
private communication) using model atmospheres of \citet{kurucz93} to
these measured equivalent widths. The spectrum synthesis technique was
applied to the absorption lines of some neutron-capture elements where
blending with other features or hyperfine splitting significantly
affects the apparent line widths. 
 
We adopted the effective temperatures determined by \citet{sneden97}
in our analysis. Surface gravities ($g$) were determined from the
ionization balance between Fe {\small I} and Fe {\small II}, while the
micro-turbulence ($v_{\rm turb}$) was determined from the Fe {\small I}
lines by demanding that there be no dependence of the derived abundance on
the equivalent widths. 

We selected lines that are detected in all of the stars in our
sample (Table~\ref{tab:ew}) to obtain final  abundances of neutron-capture
elements.   Since the \ion{Ba}{2} resonance lines at 4554
and 4934~{\AA} are very strong and inappropriate for the abundance
measurements, we adopted the equivalent widths of the three lines in
the red spectral range measured by \citet{sneden97} to determine the
Ba abundances. We excluded the \ion{Eu}{2} line at 3819~{\AA} because
of severe blending with other features. We adopted the lines at 4129
and 4205~{\AA} measured for our spectra, and the equivalent widths of
the 6645~{\AA} line provided by \citet{sneden97}. In the analyses of the 
\ion{Ba}{2}, \ion{La}{2} and \ion{Eu}{2} lines, the effects of
hyperfine splitting were included \citep[see ][for more
details]{honda04}. Though the Sr abundance is determined from the
\ion{Sr}{2} 4077~{\AA} line, we do not include its abundance in the
following discussion because an abundance deduced from only a  single line
is quite uncertain even though it is a strong line.

We estimated the random error from the standard
deviation of the abundances determined from the individual lines for each
element in each star. Since the data quality of our M15 sample is
homogeneous and the abundances were determined using identical
line sets, we adopted the average of these standard deviations of the
six M15 stars as a measure of the random error for each element.  These are
 given in
Table~\ref{tab:res}. Systematic errors of $\lesssim 0.1$~dex due to
the uncertainties of atmospheric parameters were estimated from the
analyses by changing the parameters \citep[see also][]{sneden97}. 
However, the abundance ratios of
neutron-capture elements determined from species with the same
ionization stage are insensitive to  changes in the  atmospheric
parameters \citep{honda04}. Hence, we neglect these systematic errors
when discussing  abundance {\it ratios} between
neutron-capture elements (e.g., [Y/Eu]).  However,  they are included when
the abundances of neutron-capture elements (e.g., [Eu/H]) are discussed.


We found excellent agreement between our results for the Fe and Eu
abundances and those deduced by \citet{sneden97} for the six M15
stars, given the solar abundances adopted in their work. The detailed
abundance measurements of neutron-capture elements were reported by
\citet{sneden00b} for K462.  We found some systematic differences
(~0.4~dex) in the $\log \epsilon$ values between our results and
theirs, which are not explained by the small differences of
atmospheric parameters adopted by the two studies. Since the line list
used for their analyses is not available, we could not investigate the
reason for the discrepancy. However, the differences of the abundance
{\it ratios} of Y/Eu, Zr/Eu and La/Eu, which are discussed in the next
section, are 0.01, 0.07, and 0.14~dex, respectively. We here conclude
that there is no significant discrepancy in the abundance ratios
between our work and previous studies by Sneden et al. We note that
the quality of our spectra for the blue range is higher than those of
\citet{sneden00b}, and our study used the updated line list of La by
\citet{lawler01}.

\section{Discussion}

\begin{figure} 
\includegraphics[width=8cm]{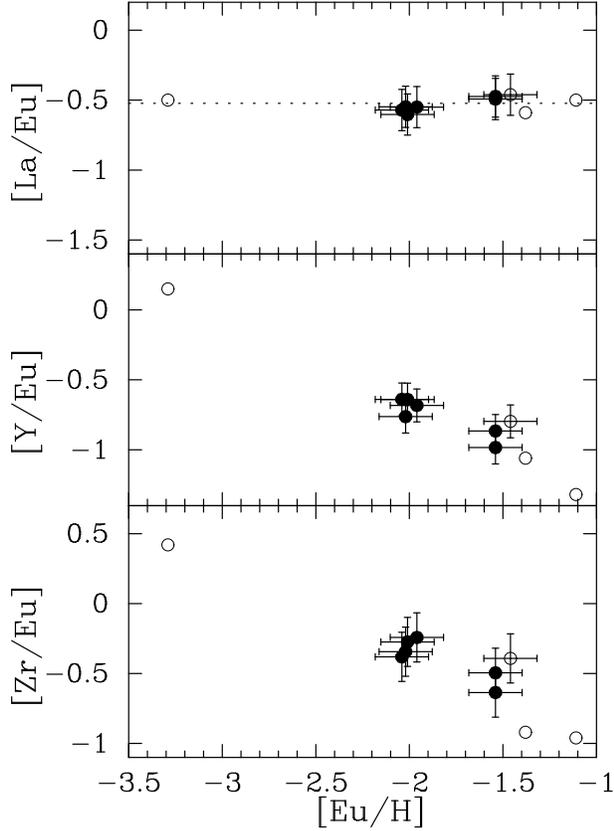}
\caption[]{Abundance ratios between neutron-capture elements as a
function of the Eu abundance ([Eu/H]). Filled circles are the values
for M15 stars obtained in the present work, while the open circles
indicate abundances of field halo stars (A symbol with error bars indicates our
results for HD221170, while the other three are from Honda et al. 2004; 
the star with a low Eu abundance is
HD~122563, and the others are the r-process enhanced stars
CS~31082--001 and CS~22892--052). The dotted line drawn on the top
panel is the La/Eu ratio of the solar-system r-process component
\citep{simmerer04}.}

\label{fig:ratios} 
\end{figure}

Figure~\ref{fig:ratios} shows the abundance ratios of neutron-capture
elements as a function of the Eu abundance, an indicator of the enrichment
by the main r-process. For comparison purposes, the values of HD~221170
and results of previous studies for r-process enhanced stars
(CS~31082--001 and CS~22892--052) and the well studied star HD~122563,
which has a relatively low Eu abundance, are also shown. 
We note that, although there are
several studies for these stars
\citep[e.g.,][]{hill02,sneden03,westin00}, we here present the results
of \citet{honda04,honda06} who adopted
the same technique and model atmosphere grid in their analysis.

There are two stars of M15 with higher abundances of neutron-capture elements
of [Eu/H]$\sim -1.5$, while others have lower values of
[Eu/H]$\sim-2.0$.  Nevertheless, the abundance ratios of La/Eu (and
Ba/Eu) are constant within the uncertainties (see the top panel of
Fig.~2).  These values are in good agreement with those of halo stars
as well as those of the r-process component in solar-system material.
These results confirm the conclusion of \citet{sneden97} that (1)
there are star-to-star abundance variations in heavy neutron-capture
elements, and (2) the heavy neutron-capture elements in M15 have
primarily originated from the r-process. Note that we deliberately
selected stars with relatively high and low Ba abundances from the
list of \citet{sneden97}.  At this point, it is still unclear as to
whether the r-process elemental abundances in M15 vary in a bimodal or
continuous fashion.


In contrast to the La/Eu ratio, the ratios of Y/Eu and Zr/Eu,
representing the abundance ratios of light to heavy neutron-capture
elements show clear anti-correlations with the Eu abundance. The
difference of the averages of the [Y/Eu] values for the four stars
with low [Eu/H]($\sim -2$) and for the other two ([Eu/H]$\sim -1.5$) is
0.34~dex, which is three times larger than the measurement error of
the Y/Eu ratio. (The differences of the averages of [Zr/Eu] values is
0.26~dex, which is only 1.5 times of the measurement error.) The
abundance ratios produced by the main r-process are estimated from the
values of the r-process enhanced halo stars. The ratios in M15 stars
show excesses of the light neutron-capture elements with respect to
the values found in r-process enriched field stars, and the excesses
are larger in stars with lower Eu abundances.

As in the context of field halo stars \citep[e.g. ][]{aoki05}, an
excess of light neutron capture elements can be explained by assuming
two sources of neutron-capture elements: one provided light
neutron-capture elements, and the other produced both light and heavy
ones (the so-called main r-process).  An example of field halo stars
representing the result of the former process is HD~122563, whose
abundance ratios are shown in Figure~\ref{fig:ratios}. The
characteristics of this process have been studied in previous works
\citep[e.g.,][]{truran02,travaglio04}, though the mechanism and the
site are not yet identified. Our results indicate that 
such a source of light neutron-capture elements operated in M15 or its
progenitor.  This source  enriched the light neutron-capture elements in the
progenitor cloud of M15 almost uniformly.  Subsequently, both
light and heavy elements formed  that were insufficiently mixed in the progenitor
proto-cluster environment.

There are two 
possibilities which explain our results.
If we assumed a simple correlation between time and the degree of mixing,
this scatter can be realized if light neutron capture elements enriched 
the progenitor of M15 earlier than the heavier elements which were not
mixed completely before star formation. 
Although the existence of (at least) two
sources of neutron-capture elements was suggested by previous studies
of field halo stars as mentioned above, no clear constraint on the time
scales of the two processes has been given (Aoki et al. 2005). Our
result that heavy neutron-capture elements show a larger scatter in their
abundances than light ones in M15 suggests that the enrichment of light
neutron-capture elements occurred in advance of the pollution of heavy ones.
This implies that the astrophysical origin of the light neutron capture elements 
is related to more massive stars than the progenitors  of the heavy r-process elements 
because heavier stars have a shorter lifetime.
Alternatively, our results could also be explained if heavy r-process 
elements are dispersed less than light ones, for example, if heavy r-process
elements were concentrated in a jet.

In either case, the dynamics of mixing of supernova ejecta and the ISM
plays a critical role.  
Unfortunately, we know little about mixing in the early Galaxy.
Our results suggest that the dynamics of mixing is  
non-negligible even in globular-cluster
progenitor clouds, although it is usually neglected on this scale
in current galactic chemical evolution models.
Few measurements of light neutron-capture elements in globular clusters
have been reported to date. 
It would be worthwhile to study the uniformity of the r-process abundance distribution in other metal-poor globular clusters to clarify the onset of mixing in the proto-cluster environment.

\acknowledgments
This work is supported at the University of Chicago in part by the
National Science foundation under Grant PHY 02-16783 for the Physics Frontier center `` Joint Institute for Nuclear Astrophysics(JINA)''.  Work at the University of Notre Dame was supported
by the U.S. Department of Energy under 
Nuclear Theory Grant DE-FG02-95-ER40934.

\clearpage

\clearpage

\clearpage

\clearpage

\begin{deluxetable}{lccccccccccc}
\tablewidth{0pt}
\tablecaption{EQUIVALENT WIDTHS OF LINES OF NEUTRON-CAPTURE ELEMENTS \label{tab:ew}}
\tablehead{ Species & Wavelength  & $\chi$ & $\log gf$   &\multicolumn{7}{c}{equivalent widths (m{\AA})\tablenotemark{a}} &Ref.\\
\cline{5-11} 
                    &  ({\AA})    &  (eV)  &      & K146 & K386 & K462 & K490 & K634 & K1040 & HD221170 &
}
\startdata
\ion{Sr}{2}& 4077.72 & 0.000  & $ 0.150$  & 175.7{$\ast$} & 252.6{$\ast$} & 220.6{$\ast$} & 175.1{$\ast$} & 204.3{$\ast$} & 198.3{$\ast$} & 277.2{$\ast$} &1\\
\ion{Y}{2} & 4854.87 & 0.990  & $-0.380$  & 26.0 & 40.7 & 50.8 & 30.9 & 38.6 & 43.2 & 45.5 &2\\
\ion{Y}{2} & 4883.68 & 1.080  & $ 0.070$  & 44.7 & 55.4 & 72.1 & 47.3 & 58.7 & 55.1 & 65.2 &1\\
\ion{Y}{2} & 5087.42 & 1.080  & $-0.170$  & 29.8 & 44.4 & 55.5 & 33.3 & 44.3 & 44.8 & 48.5 &1\\
\ion{Y}{2} & 5123.21 & 0.990  & $-0.830$  & 11.0{$\ast$} & 19.8{$\ast$} & 26.1{$\ast$} & 14.9{$\ast$} & 20.5{$\ast$} & 21.6{$\ast$} & 24.3{$\ast$} &3 \\
\ion{Y}{2} & 5200.41 & 0.990  & $-0.570$  & 20.2 & 33.0 & 47.2 & 23.7 & 32.9 & 35.1 & 39.1 &2\\
\ion{Zr}{2}& 3998.97 & 0.560  & $-0.670$  & 55.1 & 64.4{$\ast$} & 64.5{$\ast$} & 33.9{$\ast$} & 52.9{$\ast$} & 57.6{$\ast$} & 64.3 &4\\
\ion{Zr}{2}& 4050.33 & 0.710  & $-1.000$  & 19.7 & 37.3 & 40.6 & 21.1 & 30.1 & 32.1 & 37.1 &2\\
\ion{Zr}{2}& 4161.21 & 0.710  & $-0.720$  & 50.0 & 56.9{$\ast$} & 66.6{$\ast$} & 38.2{$\ast$} & 52.9{$\ast$} & 61.4{$\ast$} & 58.4{$\ast$} &1\\
\ion{Zr}{2}& 4208.98 & 0.710  & $-0.460$  & 53.1 & 71.4 & 77.5 & 47.4 & 69.9 & 70.3 & 70.7 &1\\
\ion{Zr}{2}& 4317.30 & 0.710  & $-1.380$  & 12.6 & 20.8 & 34.4 & 22.4 & 29.0 & 25.9 & 25.5 &1\\
\ion{Ba}{2}\tablenotemark{b}& 4554.03 & 0.000  & $ 0.170$  & 179.3 & 223.0 & 240.4 & 169.9 & 203.6 & 211.9 & 198.0 &5\\
\ion{Ba}{2}\tablenotemark{b}& 4934.10 & 0.000  & $-0.150$  & 180.9 & 223.7 & 235.8 & 177.1 & 206.8 & 212.7 & 208.1 &5\\
\ion{Ba}{2}\tablenotemark{c}& 5853.69 & 0.604  & $-1.010$  & 61.0 & 80.0 & 94.0 & 60.0 & 76.0 & 89.0 & 90.7 &5\\
\ion{Ba}{2}\tablenotemark{c}& 6141.73 & 0.704  & $-0.070$  & 112.0 & 133.0 & 145.0 & 108.0 & 120.0 & 143.0 & 137.0 &5\\
\ion{Ba}{2}\tablenotemark{c}& 6496.91 & 0.604  & $-0.380$  & 110.0 & 129.0 & 148.0 & 113.0 & 133.0 & 139.0 & 137.4 &8,4\\
\ion{La}{2}& 3988.52 & 0.400  & $ 0.210$  & 35.5{$\ast$} & 67.1{$\ast$} & 94.4{$\ast$} & 37.5{$\ast$} & 52.3{$\ast$} & 80.1{$\ast$} & 83.3{$\ast$} &6\\
\ion{La}{2}& 3995.75 & 0.170  & $-0.060$  & 33.0{$\ast$} & 53.4{$\ast$} & 79.8{$\ast$} & 34.6{$\ast$} & 49.6{$\ast$} & 66.6{$\ast$} & 72.3{$\ast$} &6\\
\ion{La}{2}& 4086.71 & 0.000  & $-0.070$  & 44.8{$\ast$} & 58.9{$\ast$} & 87.9{$\ast$} & 40.1{$\ast$} & 50.8{$\ast$} & 71.0{$\ast$} & 64.5{$\ast$} &6\\
\ion{La}{2}& 4123.23 & 0.320  & $ 0.130$  & 36.2{$\ast$} & 46.9{$\ast$} & 76.8{$\ast$} & 34.4{$\ast$} & 43.1{$\ast$} & 64.4{$\ast$} & 54.3{$\ast$} &6\\
\ion{La}{2}& 4333.76 & 0.170  & $-0.060$  & 48.6{$\ast$} & 63.1{$\ast$} & 93.0{$\ast$} & 51.5{$\ast$} & 59.4{$\ast$} & 81.9{$\ast$} & 85.8{$\ast$} &6\\
\ion{La}{2}& 5123.01 & 0.320  & $-0.850$  &  9.5{$\ast$} & 16.4{$\ast$} & 31.4{$\ast$} & 12.0{$\ast$} & 12.5{$\ast$} & 22.0{$\ast$} & 23.3{$\ast$} &6\\
\ion{Eu}{2}\tablenotemark{b}& 3819.67 & 0.000  & $ 0.510$  & 122.0{$\ast$} & 137.6{$\ast$} & 169.8{$\ast$} & 90.6{$\ast$} & 134.4{$\ast$} & 159.0{$\ast$} & 177.5{$\ast$} &7\\
\ion{Eu}{2}& 4129.70 & 0.000  & $ 0.220$  & 101.9{$\ast$} &  137.7{$\ast$} & 178.5{$\ast$} & 104.7{$\ast$} & 124.7{$\ast$} & 148.3{$\ast$} & 156.0{$\ast$} &7\\
\ion{Eu}{2}& 4205.05 & 0.000  & $ 0.210$  & 114.0{$\ast$} & 166.2{$\ast$} & 224.7{$\ast$} & 104.5{$\ast$} & 154.1{$\ast$} & 193.8{$\ast$} & 201.0{$\ast$} &7\\
\ion{Eu}{2}\tablenotemark{c}& 6645.13 & 1.380  & $ 0.120$  & 5.0 & 12.0 & 25.0 & 8.0 & 10.0 & 13.0 & 15.6 &7\\
\enddata
\tablenotetext{a}{Asterisks indicate synthesized values calculated for the abundance derived by spectrum synthesis.}
\tablenotetext{b}{{}Lines that were not used to derive final results.}
\tablenotetext{c}{Equivalent widths of M~15 stars are adopted from Sneden et al. (1997).}
\tablenotetext{d}{REFERENCES.-- 1. Honda et al. 2004; 2. Johnson \& Bolte 2004; 3. Hill et al. 2002; 4. Sneden et al. 2003; 5. McWilliam 1998; 6. Lawler et al. 2001a; 7. Lawler et al. 2001b; 8. Rutten 1978}
\end{deluxetable}

\clearpage

\begin{deluxetable}{lcccccccccc}
\tablewidth{0pt}
\tablecaption{ATMOSPHERIC PARAMETERS AND DEDUCED CHEMICAL ABUNDANCES \label{tab:res}}
\tablehead{ Species & $n$ & $\sigma$ & \multicolumn{8}{c}{$\log\epsilon$(X)} \\
\cline{4-10}
                    &     &          & Sun\tablenotemark{a} & K146 & K386 & K462 & K490 & K634 & K1040 & HD221170
}
\startdata
$T_{\rm eff}$ (K)   &     &          & & 4450 & 4200 & 4225 & 4350 & 4225 & 4450 & 4475 \\
$\log g$            &     &          & & 1.25 & 0.35 & 0.50 & 1.00 & 0.60 & 1.20 & 1.00 \\
$v_{\rm turb}$      &     &          & & 2.00 & 2.25 & 2.25 & 2.05 & 2.05 & 2.40 & 1.70 \\
Fe (\ion{Fe}{1})    & 54--81 & 0.133 & 7.45 & 5.12 & 5.05 & 5.10 & 5.09 & 5.20 & 5.15 & 5.41 \\ 
Fe (\ion{Fe}{2})    & 9--13  & 0.148 & 7.45 & 5.11 & 5.04 & 5.11 & 5.09 & 5.21 & 5.15 & 5.41 \\
Sr                  & 1   &         & 2.92 & 0.11 & 0.20 & 0.16 & $-0.14$ & 0.09 & 0.26 & 0.55 \\ 
Y                   & 5   &  0.041  & 2.21 & $-0.43$ & $-0.57$ & $-0.31$ & $-0.47$ & $-0.44$ & $-0.20$ & $-0.06$ \\
Zr                  & 5   &  0.137  & 2.59 & $ 0.39$ & $ 0.23$ & $ 0.41$ & $ 0.17$ & $ 0.31$ & $ 0.56$ & $ 0.74$  \\ 
Ba                  & 3   &  0.060  & 2.17 & $-0.34$ & $-0.60$ & $-0.31$ & $-0.53$ & $-0.45$ & $-0.11$ & $ 0.33$ \\
La                  & 6   &  0.098  & 1.13 & $-1.25$ & $-1.43$ & $-0.96$ & $-1.38$ & $-1.41$ & $-0.84$ & $-0.79$ \\ 
Eu                  & 3   &  0.110  & 0.52 & $-1.44$ & $-1.50$ & $-1.02$ & $-1.52$ & $-1.49$ & $-1.02$ & $-0.94$  \\ 
\enddata
\tablenotetext{a}{The solar abundances taken from \citet{asplund05}}
\end{deluxetable}

\end{document}